\documentclass[prb,twocolumn,amsmath,amssymb,superscriptaddress]{revtex4}
\usepackage[dvips]{graphicx}
\usepackage{dcolumn}
\usepackage{bm}
\usepackage{epstopdf}

\begin{document}
\title{Influence of the quantum zero-point motion of a vortex\\ on the electronic spectra of $s$-wave superconductors}

\author{Lorenz Bartosch}
\affiliation{Department of Physics, Harvard University, Cambridge,
MA 02138, USA}

\affiliation{Institut f\"ur Theoretische Physik, Universit\"at
Frankfurt, 60054 Frankfurt, Germany}

\author{Subir Sachdev}
\affiliation{Department of Physics, Harvard University, Cambridge,
MA 02138, USA}

\begin{abstract}
We compute the influence of the quantum zero-point motion of a
vortex on the electronic quasiparticle spectra of $s$-wave
superconductors. The vortex is assumed to be pinned by a harmonic
potential, and its coupling to the quasiparticles is computed in the
framework of BCS theory. Near the core of the vortex, the motion
leads to a shift of spectral weight away from the chemical
potential, and thereby reduces the zero bias conductance peak;
additional structure at the frequency of the harmonic trap is also
observed.
\end{abstract}

\pacs{pacs here}

\date{April 3, 2006}




\maketitle

\section{Introduction}

Recent work on `deconfined quantum critical points'\cite{senthil}
between superfluid and insulating phases has pointed out the central
role played by the quantum fluctuations of the vortex excitations of
the superfluid state.\cite{BBBSS05a,BBBSS05b,BBBSS05c,BBS06}
However, this work has largely been carried out using
convenient model Hamiltonians, with little direct connection to the
microscopic Bardeen-Cooper-Schrieffer (BCS) theory of the
superconducting state. It is the purpose of this paper to begin an
exploration of the quantum fluctuations of vortices in the BCS
theory. We will examine the theory of a single vortex in an $s$-wave
superconductor, fluctuating harmonically in a trapping potential.
Our primary focus will be on the influence of this quantum motion on
the electronic local density of states (LDOS). Modern scanning
tunneling microscopy (STM) can provide detailed measurements of the
LDOS in the vicinity of a vortex, and so can, in principle, detect
signatures of vortex fluctuations.

Existing $s$-wave superconductors have a large coherence length,
$\xi$, and consequently a large effective mass which is expected to
scale as the cross-section area of the vortex core $\sim \xi^2$. As
a result, the vortex zero-point motion is small, and its signatures
in the LDOS are probably indetectable. However, should a small
coherence length $s$-wave superconductor be discovered in the
future, the results here should be useful.

We will also develop some basic formalism for application to
$d$-wave superconductors. Naturally, the cuprate superconductors are
prime candidates for observing vortex zero-point motion with their
small values of $\xi$, and also indications of proximity to a
superconductor-insulator transition (possibly of the `deconfined' or
`Landau-forbidden' variety) at finite doping. However, the
application of the formalism of the present paper to $d$-wave
superconductors is far from straightforward: the anisotropic gap and
loss of rotational invariance makes the numerics much more
demanding, and the presence of gapless nodal quasiparticles leads to
new physics. We therefore defer full discussion of the $d$-wave case
to a second paper\cite{paper2} (hereafter referred to as II).

In a general sense, our analysis can be viewed as a study of the
influence of the phase of the superconducting order parameter on the
vortex LDOS. However, rather than integrating over the phase
fluctuations explicitly, we are encapsulating them in a collective
co-ordinate, the position of the vortex.

An important limitation of the calculation presented here is that,
in determining the coupling between the moving vortex and the
electronic excitations, we expand in gradients of the pairing
amplitude of the electrons. The applicability of such an expansion
places a lower bound on the value of the coherence length,
and so we are not able to properly explore the limit of extremely
short coherence lengths.
A separate computation which does not make this
gradient expansion, while working in an effective low energy theory,
will be presented in paper II.

We will begin in Section~\ref{sec:gen} by setting up the general
formalism which couples a moving vortex to the electronic
quasiparticles. Rather than considering a vortex lattice and its
oscillations, we will use a simple Einstein model for the vortex
lattice phonons, and so work with a single vortex moving in a
harmonic potential. The electronic self-energy corrections from the
vortex motion will be computed in Section~\ref{sec:swave} for an
$s$-wave superconductor. The rotation symmetry of this case is an
important aide in our numerical computations, and we are able to
obtain good numerical convergence for a significant range of
parameters.

\section{General formalism}
\label{sec:gen}

We will develop the general formalism for a two dimensional $s$- or
$d$-wave superconductor. In principle, similar considerations apply
to three-dimensional superconductors, but we will not present them
here. Complete numerical results for $s$-wave superconductors appear
in the following section, while those for $d$-wave superconductors
appear in paper II.

To model the zero-point fluctuations of a vortex in a superconductor
with either $s$- or $d$-wave symmetry we use the following
Bogoliubov-de Gennes like action as a starting point:
\begin{align}
        \mathcal{S} = &  \int d^2 r\, d\tau\,(\bar \psi_{\uparrow},\psi_{\downarrow})
        \left(\partial_{\tau} +         \mathcal{H}_{\text{BdG}} \right)
        \left(\begin{array}{c}
        \psi_{\uparrow} \\
        \bar \psi_{\downarrow}
        \end{array} \right) \nonumber \\
        & + \int d\tau\, V({\bf R}(\tau)) \;,
        \label{eq:Sinit}
\end{align}
where (with $\hbar = 1$)
\begin{equation}
\mathcal{H}_{\text{BdG}} = \left(
          \begin{array}{cc}
      \left(-\frac{1}{2 m_e} \partial_{\bf r}^2 - E_F \right) & \hat \Delta\left({\bf r} - {\bf R}(\tau) \right) \\
      \hat \Delta^{\ast}\left({\bf r} - {\bf R}(\tau) \right) & - \left(-\frac{1}{2 m_e} \partial_{\bf r}^2  - E_F \right)
    \end{array}
        \right) \;.
        \label{eq:HBdG}
\end{equation}
Here $\psi({\bf r},\tau)$ and $\bar \psi({\bf r},\tau)$ are
conjugate Grassman fields for electrons with mass $m_e$, $E_F =
k_F^2/2m_e = m_e v_F^2/2$ is the Fermi energy and $\hat \Delta({\bf
r} - {\bf R}(\tau))$ is the gap operator for superconducting
electrons in the presence of a vortex at position ${\bf R}(\tau)$ in
the plane. The vortex is allowed to move in imaginary time $\tau$
and to account for the interaction between vortices in a vortex
lattice we put our vortex in a harmonic oscillator potential $V({\bf
R})$ with equilibrium position ${\bf R} = 0$.

For a superconductor with $s$-wave symmetry the gap operator can be
taken to be a simple scalar $\Delta({\bf r})$. Assuming there is a
vortex at the center of the plane we can choose a gauge such that
the phase of the order parameter is equal to the polar angle of the
position vector, i.e. (with a slight abuse of notation) $\Delta({\bf
r}) = \Delta(r) e^{i\theta}$.

The $d$-wave case is more difficult and it is customary to express
the gap operator in terms of a non-local order parameter with a
center of mass coordinate ${\bf r}$ and a relative coordinate ${\bf
r}'$. Expanding in powers of ${\bf r}'$ and keeping only terms up to
second order we can eliminate ${\bf r}'$ and write the gap operator
with $d_{x^2 - y^2}$ symmetry (see Appendix \ref{Appendix:gauge}) as
\begin{equation}
  \hat\Delta = \frac{\{\partial_x,\{\partial_x,\Delta({\bf r})\}\}}{k_F^2}
  - \frac{\{\partial_y,\{\partial_y,\Delta({\bf r})\}\}}{k_F^2} \;,
  \label{eq:DeltaSL}
\end{equation}
where $\{a,b\}=(ab+ba)/2$. This equation is equivalent to a gap
operator with $d_{xy}$ symmetry as derived by Simon and
Lee.\cite{Simon97} For a vortex at the origin we again take
$\Delta({\bf r}) = \Delta(r) e^{i\theta}$.
In Appendix~\ref{Appendix:gauge} we will discuss issues of
gauge invariance associated with the above expressions and the important
comments made on
this issue in Ref.~\onlinecite{Vafek01}. A choice of gauge is made
in Eq.~(\ref{eq:DeltaSL}), and no additional term is needed to
preserve gauge invariance of the Bogoliubov-de Gennes equations.

Integrating out the electronic degrees of freedom in the action
given in Eq.~(\ref{eq:Sinit}) one obtains an effective action for
the vortex degrees of freedom. Expanding in powers of the velocity
of the vortex we obtain for the effective vortex action (for
simplicity at zero temperature) in the imaginary frequency formalism
(see Appendix \ref{app:vortexaction})
\begin{equation}
  \mathcal{S}_{\text{eff}}^{\text{Vortex}} = \frac{m_v}{2} \int \frac{d \omega}{2\pi}\,
  {\bf R}^{\dagger} (i\omega)
  \left( \begin{array}{cc}
      \omega^2 + \omega_0^2 & \omega_c \omega \\
      -  \omega_c \omega & \omega^2 + \omega_0^2
    \end{array} \right) {\bf R} (i\omega) \;.
  \label{eq:Vortex}
\end{equation}
While the term proportional to $\omega$ can be identified to be the
Magnus force, the term proportional to $\omega^2$ is just the vortex
kinetic energy and defines the mass of the vortex, $m_v$. For a
BCS-superconductor with $s$-wave symmetry $m_v$ is of the order of
$m_e (k_F \xi)^2$. With $m_v$ given, we have defined $\omega_0$ by
$V({\bf R}) = m_v \omega_0^2 {\bf R}^2 /2$. For a vortex in a vortex
lattice, the characteristic frequency is the plasma frequency
\begin{equation}
\omega_p = \sqrt{4\pi^2 \rho_s/m_v A_0} \;,
\end{equation}
where $\rho_s$ is the superfluid stiffness and $1/A_0$ is the
density of vortices. The frequency $\omega_0$ is approximately given
by $\omega_0 \approx (5/2) \omega_p$.\cite{BBS06} Finally, in a
Galilean invariant superfluid, $\omega_c = 2\pi \rho_s/m_v$ and in a
dual picture can be identified to be the `cyclotron' frequency. For
a superfluid on a lattice, however, it has recently been argued that
$\omega_c$ is reduced and the density of the Mott insulator has to
be subtracted from the superfluid stiffness when the superfluid is
close to a nearby Mott insulating
state.\cite{BBBSS05a,BBBSS05b,BBBSS05c,BBS06}

Diagrammatically, the above propagator to the effective vortex
action is just the RPA propagator which includes the Berry phase
term. To determine the effect of the vortex motion on the electronic
spectrum we can use the RPA propagator and calculate the self energy
correction to the electronic eigenstates in the
GW approximation.\cite{Hedin65}
Let us first expand the gap operator $\hat \Delta \left({\bf r} -
{\bf R}(\tau) \right)$ in Eq.~(\ref{eq:Sinit}) to leading order in
${\bf R(\tau)}$. Then $\mathcal{S}$ can be written as $\mathcal{S}
\approx \mathcal{S}_0 + \mathcal{S}_{int}$ with
\begin{equation}
        \mathcal{S}_0 = \int d\tau\, d^2 r\, (\bar \psi_{\uparrow},\psi_{\downarrow})
        \left(\partial_{\tau} +         \mathcal{H}_{\text{BdG}}^0 \right)
        \left(\begin{array}{c}
        \psi_{\uparrow} \\
        \bar \psi_{\downarrow}
        \end{array} \right)
        \label{eq:S0}
\end{equation}
and $\mathcal{H}_{\text{BdG}}^0$ obtained from Eq.~({\ref{eq:HBdG})
by setting ${\bf R} =0$. To leading order the coupling of the vortex
to the electronic degrees of freedom is described by
\begin{equation}
        \mathcal{S}_{int} = - \int d\tau\, d^2 r \,
        {\bf R(\tau)} \cdot (\bar \psi_{\uparrow},\psi_{\downarrow})
        \left(\begin{array}{cc}
        0 & {\bf \partial}_{\bf r} \hat \Delta \\
        {\bf \partial}_{\bf r} \hat \Delta^{\ast} & 0
        \end{array} \right)
        \left(\begin{array}{c}
        \psi_{\uparrow} \\
        \bar \psi_{\downarrow}
        \end{array} \right) \;.
\end{equation}
Let us now go to a basis which diagonalizes the Bogoliubov-de Gennes
Hamiltonian $\mathcal{H}_{\text{BdG}}^0$: It is well-known that if
$\Psi_{\ell}({\bf r}) \equiv [u_\ell({\bf r}),v_\ell({\bf r})]^T$
(with, say, $\ell > 0$) is an eigenstate of
$\mathcal{H}_{\text{BdG}}^0$ with eigenvalue $\epsilon_\ell$, then
$[-v_\ell^{\ast}({\bf r}),u_\ell^{\ast}({\bf r})]^T$ is an
eigenstate of $\mathcal{H}_{\text{BdG}}^0$ with eigenvalue
$-\epsilon_\ell$.\cite{deGennes:Book} Defining
\begin{equation}
        U_\ell({\bf r}) = \left(\begin{array}{cc}
        u_\ell({\bf r}) & -v_\ell^{\ast}({\bf r}) \\
        v_\ell({\bf r}) & u_\ell^{\ast}({\bf r})
        \end{array} \right)
\end{equation}
we can write
\begin{equation}
        \left(\begin{array}{c}
        \psi_{\uparrow} ({\bf r},\tau)\\
        \bar \psi_{\downarrow} ({\bf r},\tau)
        \end{array} \right)
        = \sum_{\ell>0} U_{\ell}({\bf r})
        \left(\begin{array}{c}
        \chi_{+\ell} (\tau)\\
        \chi_{-\ell} (\tau)
        \end{array} \right)
\end{equation}
such that
\begin{equation}
  \int d^2r \, U_\ell^{\dagger} ({\bf r}) \mathcal{H}_{\text{BdG}}^0 U_{\ell^{\prime}}({\bf r}) = \sigma_z \, \epsilon_\ell \, \delta_{\ell,\ell^{\prime}} \; ,
\end{equation}
where $\sigma_z$ (like $\sigma_x$ which we will use below) is just a usual Pauli matrix. 
While it is customary to restrict the energies $\epsilon_\ell$ to
values greater zero\cite{deGennes:Book} (for which we use quantum
numbers $\ell > 0$), we prefer to subsume the `spin' index into the
index $\ell$ and define for $\ell > 0$ $\Psi_{-\ell}({\bf r}) \equiv
[u_{-\ell}({\bf r}),v_{-\ell}({\bf r})]^{T} \equiv
[-v_\ell^{\ast}({\bf r}),u_\ell^{\ast}({\bf r})]^{T}$ such that
$\epsilon_{-\ell} = -\epsilon_\ell$. With this notation,
\begin{equation}
  \left(\begin{array}{c}
      \psi_{\uparrow} ({\bf r},\tau)\\
      \bar \psi_{\downarrow} ({\bf r},\tau)
    \end{array} \right)
  = \sum_{\ell} \Psi_{\ell}({\bf r})\, \chi_{\ell}(\tau)
      \label{eq:psitochi}
\end{equation}
and $\mathcal{S}_0$ reduces to
\begin{equation}
        \mathcal{S}_0 = \int d\tau\, \sum_\ell \bar \chi_\ell (\partial_{\tau} + \epsilon_\ell) \chi_\ell \;,
        \label{eq:S0II}
\end{equation}
where the sum is now over all quantum numbers $\ell$ including the
`spin' index.
 In the new basis, $\mathcal{S}_{int}$ can be written as
\begin{equation}
        \mathcal{S}_{int} = - \int d\tau \, \sum_{\ell \ell^{\prime}}
        {\bf R(\tau)} \cdot  {\bf M}_{\ell;\ell^{\prime}}
        \bar \chi_{\ell} \chi_{\ell^{\prime}} \;,
\end{equation}
where the transition matrix elements
\begin{equation}
        {\bf M}_{\ell;\ell^{\prime}} =
        \int d^2r \,\left( u_\ell^{\ast} \partial_{\bf r} \hat \Delta v_{\ell'} + v_\ell^{\ast} \partial_{\bf r} \hat \Delta^{\ast} u_{\ell'} \right)
\end{equation}
are like ${\bf R}(\tau)$ vectors in the two-dimensional plane. It is
convenient to also define $M^{\pm} \equiv (M^{x} \pm i M^{y})/2$
such that
\begin{equation}
        M_{\ell;\ell^{\prime}}^{+} = \left(M_{\ell^{\prime};\ell}^{-}\right)^{\ast} =
        \int d^2r \,\left( u_\ell^{\ast} \partial_{\bar z} \hat \Delta v_{\ell'} + v_\ell^{\ast} \partial_{\bar z} \hat \Delta^{\ast} u_{\ell'} \right) \;,
        \label{eq:Mplusminus}
\end{equation}
with 
$\partial_{\bar z} \equiv (\partial_x + i \partial_y)/2$. We can now
calculate the self energy in the GW approximation\cite{Hedin65} for
which we
obtain 
\begin{equation}
  \Sigma_{\ell} (i \tilde \omega) =
  \sum_{\ell^{\prime} \alpha}
  \frac{A_{\ell;\ell'}^{\alpha}} {i \tilde \omega - [\omega_v \, \text{sgn} \,(\epsilon_{\ell'}) + \epsilon_{\ell'} ]
    - \alpha \omega_c/2} \;,
\label{eq:SelfEnergy}
 \end{equation}
with
\begin{equation}
  \label{eq:A}
  A_{\ell;\ell'}^{\alpha} = \frac{ | M_{\ell;\ell^{\prime}}^{\alpha} |^2}{m_v \omega_v} \; .
\end{equation}
 Here, $\alpha = \pm$ and $\omega_v = \sqrt{\omega_0^2 + \omega_c^2/4}$ is the vortex (`magnetoplasma') frequency in an Einstein model. As should be expected, the self energy satisfies $\Sigma_{-\ell}(i\tilde\omega) = -\Sigma_{\ell}(-i\tilde\omega)$.

If our system is infinitely large and  boundary effects can be
neglected we can make use of the Hellmann-Feynman theorem
\begin{equation}
  \label{eq:HellFey}
  \int d^2 r\, \Psi_{\ell}^{\dagger} \, \partial_{\bf r} \mathcal{H}_{BdG}^0\, \Psi_{\ell'} = (\epsilon_{\ell'} - \epsilon_{\ell}) \int d^2 r\, \Psi_{\ell}^{\dagger} \partial_{\bf r} \Psi_{\ell'}
\end{equation}
and write $M_{\ell,\ell'}^{\alpha}$ as
\begin{equation}
  \label{eq:MWII}
  M_{\ell,\ell'}^{\alpha} = (\epsilon_{\ell'} - \epsilon_{\ell}) U_{\ell,\ell'}^{\alpha} \;,
\end{equation}
with
\begin{equation}
  \label{eq:def:W}
  U_{\ell,\ell'}^{+} = \left( U_{\ell',\ell}^{-} \right)^{\ast} =
  \int d^2 r\, \Psi_{\ell}^{\dagger} \partial_{\bar z} \Psi_{\ell'} \;.
\end{equation}
In our numerical calculation presented below, however, we will, of
course, consider a system of finite size such that boundary effects
need to be included. We will see that
calculating $M_{\ell,\ell'}^{\alpha}$ using the
Hellmann-Feynman theorem will turn out to be easier than
calculating these transition matrix elements directly.

\section{Vortex in an $s$-wave superconductor}
\label{sec:swave}

 For a vortex in an $s$-wave superconductor centered at the origin we choose $\Delta({\bf r}) = \Delta(r)\,e^{i\theta}$ with $\Delta(r) = \Delta_0 \tanh(r/\xi)$ where $\Delta_0$ is the bulk gap and $\xi = v_F/\pi \Delta_0$ is the coherence length.
$\mathcal{H}_{\text{BdG}}^0$ is rotationally invariant such that
angular momentum is a good quantum number. Following Caroli, de
Gennes, and Matricon, \cite{Caroli64} we denote angular momentum by
$\mu = \pm 1/2, \pm 3/2,\dots$ and write
 \begin{equation}
        \left(\begin{array}{c}
                u_\mu^n ({\bf r}) \\ v_\mu^n ({\bf r})
        \end{array}\right) =
        \frac{\exp\left[ -i(\mu - \sigma_z /2) \theta \right]}{\sqrt{2\pi}}
        \left(\begin{array}{c}
                f_{\mu,+}^n (r) \\ f_{\mu,-}^n (r)
        \end{array}\right) \;.
\label{eq:uf}
\end{equation}
There is only one bound state for each angular momentum $\mu$, but
there are also extended states, such that we also include a radial
quantum number $n$.\footnote{Actually, $\mu$ is minus the angular
momentum of a given eigenstate. The minus sign is due to the fact
that we are considering a vortex of positive vorticity but prefer to
have $\epsilon_{\mu}$ positive if $\mu$ is positive.} It should be
noted that while $\ell$ is a collective label for $\mu$ and $n$,
$-\ell$ collectively labels $-\mu$ and $n$. Making the above ansatz,
the Bogoliubov-de Gennes equations reduce to
\begin{widetext}
\begin{align}
& \left\{ \sigma_z \frac{1}{2 m_e} \left( -\left(\frac{d}{d
r}\right)^2 - \frac{1}{r} \frac{d}{dr} + \frac{(\mu -
\sigma_z/2)^2}{r^2} - k_F^2 \right)
+ \sigma_x \Delta(r) \right\} \left(
  \begin{array}{c}
    f_{\mu,+}^{n} (r) \\ f_{\mu,-}^{n} (r)
  \end{array}
\right) = \epsilon_{\mu}^n \left(
  \begin{array}{c}
    f_{\mu,+}^{n} (r) \\ f_{\mu,-}^{n} (r)
  \end{array}
\right) \;. \label{eq:HBdGf}
\end{align}
\end{widetext}

Before solving these equations numerically, it is worthwhile to
first consider the bound states with energies $\epsilon_{\mu}$ much
smaller than the bulk gap $\Delta_0$ and large coherence lengths
$\xi$. As was shown by Caroli, de Gennes, and Matricon
\cite{Caroli64}, the bound state energies are then given by
\begin{equation}
  \label{eq:epsilon_mu}
  \epsilon_{\mu} = E_1 \, \mu \;,\quad
  \text{with} \quad E_1 \approx \frac{\Delta_0^2}{E_F} \;.
\end{equation}
Also, the radial part of the wave function is well approximated by

\begin{equation}
  \label{eq:f_mu}
  f_{\mu,\pm}(r) = C_{\mu} \exp\left(
    -2 \int_0^r dr' \, \frac{\Delta(r')}{v_F} \right) J_{\mu \mp 1/2} (k_F r) \;,
\end{equation}
where $J_m(x)$ are ordinary Bessel functions of the first kind and
integer order $m=\mu \mp 1/2$. The constants $C_{\mu}$ are
independent of the $\pm$ index and are all of order
$\sqrt{k_F/\xi}$. It should be noted that $f_{\mu+1,+}(r) \approx
f_{\mu,-}(r)$ which we will use for an estimate of the matrix
elements $A_{\mu;\mu'}^{+} \equiv \delta_{\mu',\mu+1} A_{\mu}^{+}$.
With the radial quantum number included we have
\begin{widetext}
\begin{align}
  M_{\mu n;\mu' n'}^{+} = \frac{1}{2} & \delta_{\mu',\mu+1} \int_0^{\infty} dr\,
  \left\{ \left[ r \partial_r \Delta - \Delta(r) \right]
    f_{\mu,+}^{n}(r)
f_{\mu',-}^{n'}(r) + \left[ r \partial_r \Delta + \Delta(r) \right]
    f_{\mu,-}^{n}(r)
 f_{\mu',+}^{n'}(r)
\right\} \;.
\end{align}
\end{widetext}
Using the above, we obtain $A_{\mu}^{+} \approx v_F^2/(4\pi^2 m_v
\omega_v \xi^4)$. Since the mass of a vortex in a BCS
superconductor, $m_v$, is of the order $m_e (k_F \xi)^2$, it follows
\begin{equation}
  \label{eq:Amu}
  A_{\mu}^{+} \approx A \equiv \frac{1}{4\pi v_F m_e^3 (\omega_v/\Delta_0) \xi^5} \;.
\end{equation}
The fact that if we keep $\omega_v/\Delta_0$ constant, the self
energy increases with the fifth power of $1/\xi$ as $\xi$ decreases
is quite remarkable. For large $\xi$, the self energy correction is
very small and the motion of the vortex obviously has practically no
influence on the spectrum. However, as $\xi$ decreases, the self
energy correction becomes more and more important and we expect a
dramatic change of the spectrum within a small range of the
coherence length $\xi$.

Let us now consider the local density of states (LDOS) which (in the
more general case and with $\ell$ including the `spin' index) is
given by
\begin{equation}
  \label{eq:lDOS}
  \rho({\bf r},\omega) = -\frac{1}{\pi}\, \text{Im}\, \sum_{\ell}
  \frac{|u_{\ell}({\bf r})|^2}{\omega - \epsilon_{\ell} - \Sigma_{\ell}(\omega) + i0^{+}} \;.
\end{equation}
For the case of an $s$-wave order parameter as considered here, the
only bound state wave function which does not vanish at $r=0$ is
$u_{\mu=1/2}(r)$. We can therefore actually calculate the LDOS
at the vortex center and obtain
\begin{align}
  \label{eq:DOSat0}
  \rho(r&=0,\omega) = \frac{|u_{\mu=1/2}(0)|^2}{\omega_1^2} \big\{
(\omega_v + \omega_c + E_1)^2 \,
    \delta(\omega-\epsilon_{1/2}) \nonumber \\
  & + A \,
    \delta(\omega-\epsilon_{1/2} - \omega_1)
    + A \,
    \delta(\omega-\epsilon_{1/2} + \omega_1) \big\} \;,
\end{align}
with $\omega_1 = \sqrt{(\omega_v + \omega_c + E_1)^2 + 2A}$ which
for large coherence lengths $\xi$ is close to the vortex frequency
$\omega_v$. While there is just a peak with weight
$|u_{\mu=1/2}(0)|^2$ at the unperturbed energy $\epsilon_{1/2}$ in
the absence of the matrix element $A$, as $\xi$ decreases $A$
increases and we also find two satellite peaks shifted from this
position by $\pm \omega_1$.

To obtain the LDOS away from the vortex center and to calculate
the LDOS at arbitrary energy we follow Gygi and Schl\"uter
\cite{Gygi91} and evaluate  the eigenenergies and eigenvectors of
$\mathcal{H}_{\text{BdG}}^0$ numerically. First we replace the
system of infinite size by a disk of finite radius $R_0$. We then
expand the quasi-particle amplitudes $f_{\mu,\pm}^{n}$ into
Fourier-Bessel series: Denoting the $j$'th zero of the ordinary
Bessel function $J_m$ by $\alpha_{mj}$, we introduce the functions
\begin{equation}
  \label{eq:phi}
  \phi_{mj}(r) = \frac{\sqrt{2}}{R_0 |J_{m+1}(\alpha_{mj})|} J_m(\alpha_{mj}r/R_0) \;,
\end{equation}
which are eigenfunctions to the kinetic energy operator and satisfy
the normalization condition
\begin{equation}
  \label{eq:phinormalization}
  \int_0^{R_0} dr \, r \, \phi_{mj}(r) \phi_{mj'}(r) = \delta_{j j'} \;.
\end{equation}
Truncating the Fourier-Bessel series for the quasi-particle
amplitudes $f_{\mu,\pm}^n$ at large $N_0$, we have
\begin{equation}
  \label{eq:fFourierBessel}
  \left(\begin{array}{c}
      f_{\mu,+}^n (r) \\ f_{\mu,-}^n (r)
    \end{array}\right) = \sum_{j=1}^{N_0} \left(
    \begin{array}{c}
      c_{\mu j}^n \phi_{\mu + 1/2,j} (r) \\
      d_{\mu j}^n \phi_{\mu - 1/2,j} (r)
    \end{array}\right)
\end{equation}
With these approximations the Bogoliubov-de Gennes equations can be
solved by solving the following matrix eigenvalue problem:
\begin{equation}
  \label{eq:BdGMatrix}
  \left(
    \begin{array}{cc}
      T^{-} & \Delta \\
      \Delta^T & T^{+}
    \end{array}
  \right) \Psi_{\mu}^n = \epsilon_{\mu}^n \Psi_{\mu}^n \;.
\end{equation}
Here $T^{\pm}$ and $\Delta$ are $N_0 \times N_0$ matrices with
matrix elements
\begin{align}
  \label{eq:T}
  T_{j j'}^{\pm} = & \, \mp \frac{1}{2m_e} \left(
    \frac{\alpha_{\mu\pm 1/2,j}^2}{R_0^2} - k_F^2 \right) \delta_{j j'}\;,  \\
  \Delta_{j j'} = & \int_0^{R_0} dr\, r \, \phi_{\mu-1/2,j}(r)\, \Delta(r)\, \phi_{\mu+1/2,j'}(r) \;,
\end{align}
and $\Psi_{\mu}^{n}$ is given by $\Psi_{\mu}^{n} = (c_1 \cdots
c_{N_0},d_1 \cdots d_{N_0})^T$. Having calculated these eigenvectors
we can express the matrix elements $M_{\mu,n;\mu+1,n'}^{+}$ which
determine the $A_{\mu;n n'}^{+}$ as
\begin{equation}
  \label{eq:MKL}
  M_{\mu,n;\mu+1,n'}^{+} =
  \frac{1}{2} \sum_{j j'}
  \left( c_{\mu j}^n K_{j j'}^{(\mu)} d_{\mu+1,j'}^{n'} +
    d_{\mu j}^n L_{j j'}^{(\mu)} c_{\mu+1,j'}^{n'} \right) \;,
\end{equation}
where we have defined
\begin{equation}
  K_{j j'}^{(\mu)} = \int_0^{R_0} dr\, [r \partial_r \Delta - \Delta(r)] \,
  \phi_{\mu-1/2,j}(r) \,
  \phi_{\mu+3/2,j'}(r) \;,
\end{equation}
and
\begin{equation}
  L_{j j'}^{(\mu)} = \int_0^{R_0} dr\, [r \partial_r \Delta + \Delta(r)] \,
  \phi_{\mu+1/2,j}(r) \,
  \phi_{\mu+1/2,j'}(r) \;.
\end{equation}
Alternatively, as discussed at the end of the preceding section, we
can also make use of the Hellmann-Feynman theorem which when
including boundary terms reads for the $s$-wave case considered
here\cite{Han05}
\begin{equation}
  \label{eq:Hellmannswave}
  M_{\ell,\ell'}^{+} = (\epsilon_{\ell'} - \epsilon_{\ell}) U_{\ell;\ell'}^{+} - \frac{R_0}{2 m_e} \int_0^{2\pi} d\theta\, \partial_r \Psi_{\ell}^{\dagger} \sigma_3 \partial_{\bar z} \Psi_{\ell'} \Big|_{r = R_0} \;.
\end{equation}
Expressing all wave functions in terms of their Fourier-Bessel
components both integrals can be done analytically and we obtain
\begin{align}
  \label{eq:MKK}
  & \!\!\!\!M_{\mu,\mu+1}^{n n'} =  \nonumber \\
 & \frac{1}{2}
  \sum_{j j'}
  c_{\mu j}^n \left[(\epsilon_{\mu+1}^{n'} - \epsilon_{\mu}^{n}) \mathcal{K}_{j j'}^{(\mu-1/2)} - \mathcal{L}_{j j'}^{(\mu-1/2)} \right] c_{\mu+1,j'}^{n'} \nonumber \\
  + & \frac{1}{2}
  \sum_{j j'}
  d_{\mu j}^n \left[(\epsilon_{\mu+1}^{n'} - \epsilon_{\mu}^{n}) \mathcal{K}_{j j'}^{(\mu+1/2)} + \mathcal{L}_{j j'}^{(\mu+1/2)} \right] d_{\mu+1,j'}^{n'} \;,
\end{align}
with the matrix elements $\mathcal{K}_{j j'}^{(m)}$ and
$\mathcal{L}_{j j'}^{(m)}$ given by
\begin{align}
  \label{eq:def:K}
  \mathcal{K}_{j j'}^{(m)} & = \text{sign}\,(m+1/2)\,(-1)^{j-j'}\,\frac{2 \alpha_{mj} \alpha_{m+1,j'}}{R_0(\alpha_{m+1,j'}^2 - \alpha_{mj}^2)} \;, \\
  \label{eq:def:L}
  \mathcal{L}_{j j'}^{(m)} & = \text{sign}\,(m+1/2)\,(-1)^{j-j'}\,\frac{\alpha_{mj} \alpha_{m+1,j'}}{m_e R_0^3} \;.
\end{align}
For simplicity and to avoid too many parameters which do not change
the essential physics, we have neglected the electromagnetic vector
potential in our above considerations and will not calculate the
pair potential self-consistently. Neglecting the vector potential is
safe in the extreme type II case where the London penetration length
is much larger than the coherence length which we are considering
here. Although there are significant deviations from the
$\tanh$-behavior of $\Delta(r)$ for small temperatures, corrections
to the LDOS are expected to be small and can easily be
incorporated. Once all eigenenergies and eigenvectors of
$\mathcal{H}_{\text{BdG}}^0$ are determined, the matrix elements
$A_{\mu;nn'}^{+}$ and then the LDOS can be calculated.

In STM experiments the local tunneling conductance $G=\partial I/
\partial V$ can be measured as a function of gate voltage $V$. If we
are at very low temperature and have a tip with a constant density of states (DOS) the
tunneling conductance is essentially equal to the LDOS of the
probe. At finite temperature, each peak in the LDOS becomes
broadened with the derivative of the Fermi function $f(\omega) =
1/(e^{\omega/T}+1)$ such that
\begin{equation}
  \label{eq:G}
  G({\bf r},\omega=eV) = -\frac{G_0}{\rho_0} \int d\omega' \,\rho({\bf r},\omega+\omega')\,f^{\prime}(\omega') \;.
\end{equation}
Here we have expressed the normalization constant in terms of the
DOS of a free 2-dimensional electron gas (per spin direction),
$\rho_0 = m_e/2\pi$, and the corresponding tunneling conductance,
$G_0$.

In our numerical calculations we have chosen $k_F R = 400$. For the
vortex (Einstein) frequency we use $\omega_v = 0.2\, \Delta_0$ and
for simplicity we set $\omega_c=0$. Setting the temperature equal to
$T=0.006\,E_F$ we obtain smooth curves for the tunneling conductance
with well distinguished bound state peaks. In Fig.~\ref{fig:xi=10}
\begin{figure}[tb]
  \includegraphics[width=8cm]{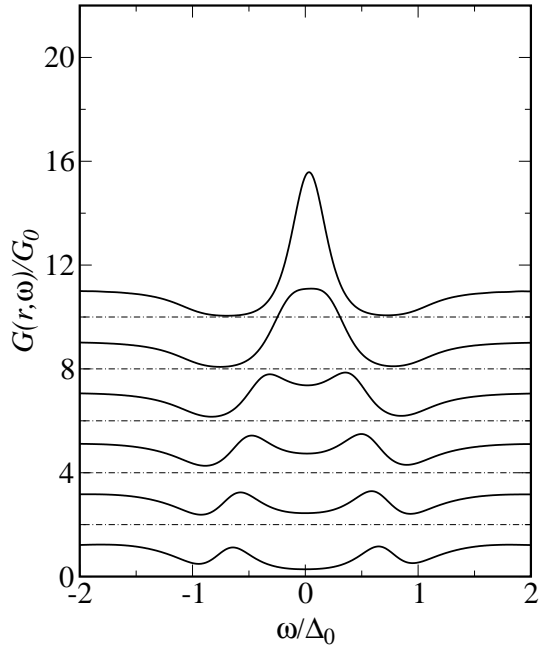}
  \caption{Tunneling conductance $G=\partial I / \partial V$ for a superconductor with
  $s$-wave symmetry as a function of $\omega$ at the vortex center $r=0$ (upper curve)
  and $k_F r = 4,8,\dots,20$ for $k_F \xi=10$ and $m_v = m_e(k_F\xi)^2 = 100\, m_e$.
  Each curve is offset by $2$ units for clarity. We have chosen $\omega_v = 0.2\, \Delta_0$
  and have for simplicity set $\omega_c=0$. The temperature is $T=0.006\, E_F$ and
  this finite temperature leads to a broadening of the conductance peaks.
  As can be seen in the figure, there are no satellite peaks. For these parameters, the tunneling conductance
  of the static vortex case, without the self energy correction, has an almost identical appearance.}
\label{fig:xi=10}
\end{figure}
\begin{figure}[tb]
  \centering
  \includegraphics[width=8cm]{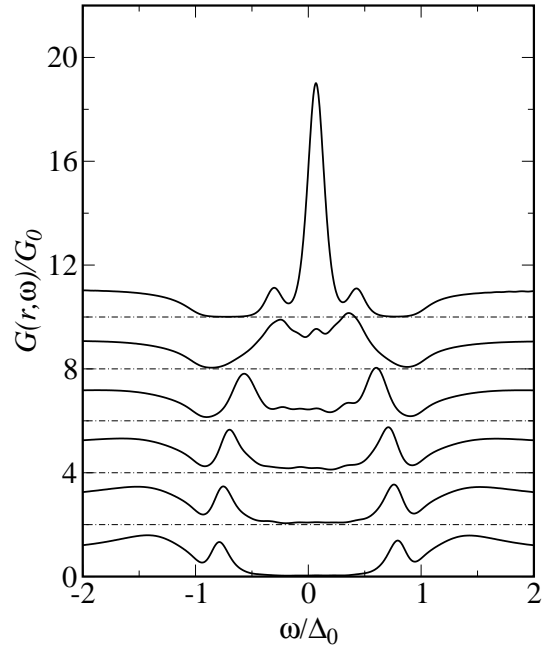}
  \caption{Tunneling conductance $G=\partial I / \partial V$ for a smaller value of $\xi$ than that
    in Fig.~\ref{fig:xi=10}, $k_F \xi=5$ and $m_v = m_e(k_F \xi)^2 = 25\, m_e$. The tunneling conductance
    is plotted as a function of $\omega$ at the
    vortex center $r=0$ (upper curve) and $k_F r = 4,8,\dots,20$.
    All other parameters are chosen as in Fig~\ref{fig:xi=10}. The two
    satellite peaks shifted from the central peak by $\pm \omega_1$ can
    clearly be seen in the curve for $r=0$.\\ \\ \\ \\ {}} \label{fig:xi=5}
\end{figure}
\begin{figure}[tb]
  \centering
  \includegraphics[width=8cm]{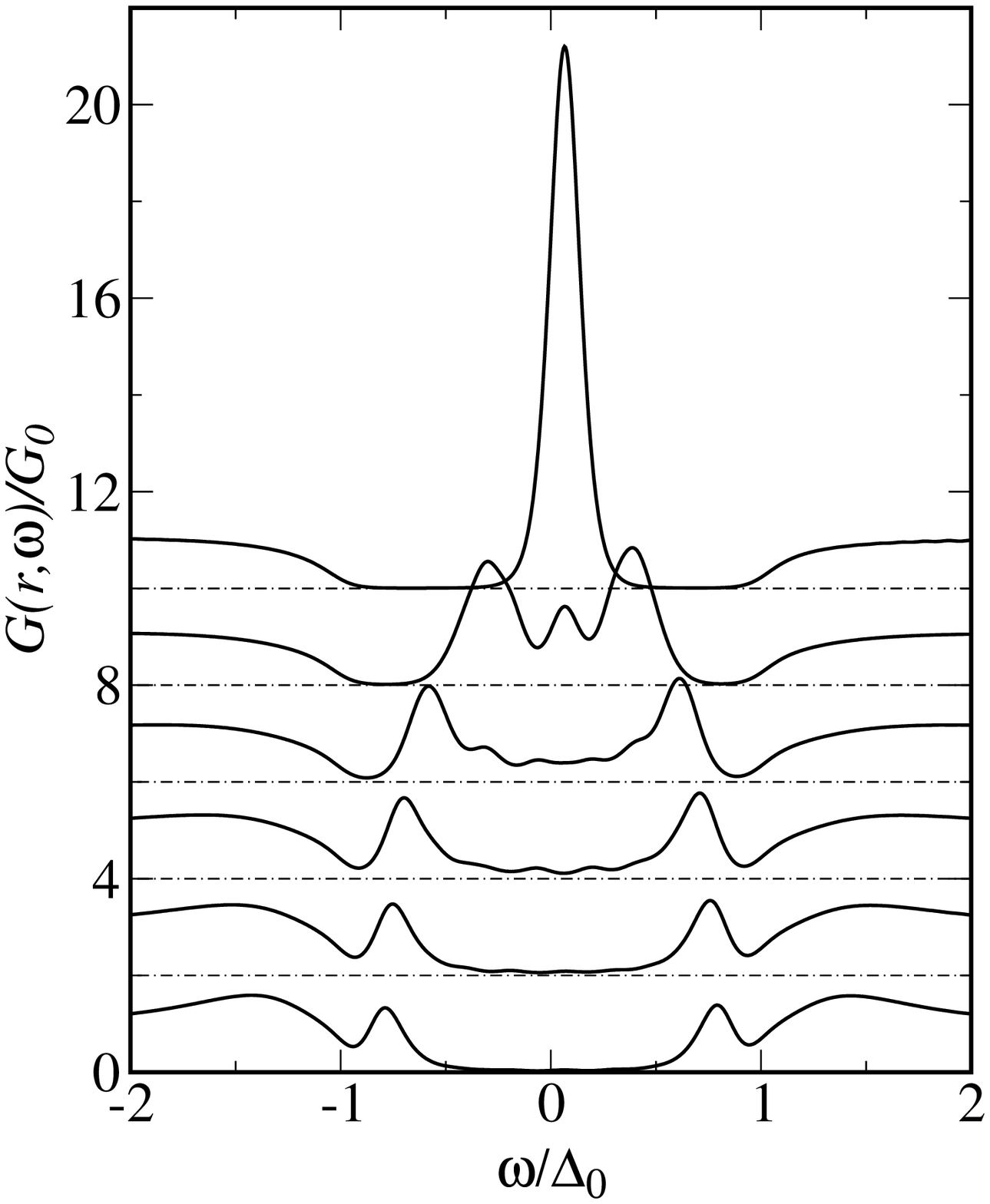}
  \caption{Tunneling conductance $G=\partial I / \partial V$ for the same parameters as in Fig.~\ref{fig:xi=5}, but with a static vortex. }
  \label{fig:xi=5:static}
\end{figure}
we show plots of the tunneling conductance $G({\bf r},\omega)$ at
the vortex center and at several distances away from it for $k_F \xi
= 10$ and $m_v = m_e (k_F \xi)^2 = 100\,m_e$. The plots are almost
indistinguishable from plots for the tunneling conductance with the
same parameters but a static vortex (this amounts to taking the
limit $m_v \to \infty$).\cite{Gygi91} As we decrease the coherence
length (and/or the vortex mass) the influence of the moving vortex
on the electronic spectrum becomes more and more pronounced. In
Fig.~\ref{fig:xi=5} we show the tunneling conductance for $k_F \xi =
5$ and $m_v = m_e(k_F \xi)^2 = 25\,m_e$. Now one can clearly see two
satellite peaks at the center of the vortex, shifted from the
central peak by $\pm \omega_1$ which is of the order of the vortex
frequency $\omega_v$. In units of the Fermi energy the temperatures
chosen in Figs.~\ref{fig:xi=10} and \ref{fig:xi=5} are the same.
However, in units of the bulk gap $\Delta_0$, the temperature in the
latter figure is smaller than that in the former by a factor of two.
As a result of this, the hight of the zero bias peak is actually
larger in Fig.~\ref{fig:xi=5} than in Fig.~\ref{fig:xi=10} even
though spectral weight is shifted to the satellite peaks in
Fig.~\ref{fig:xi=5}. In fact, the quantum zero-point motion does
lead to a reduction of the zero bias conductance peak as can be seen
by comparing the results of Fig.~\ref{fig:xi=5} with those
calculated for the same parameters, but with a static vortex. This
case is shown in Fig.~\ref{fig:xi=5:static}.

\section{Conclusions}
\label{sec:conc}

We presented a first determination of the influence of the vortex
zero-point motion in a simple model based upon BCS theory. For
$s$-wave superconductors, a comprehensive computation was possible,
with good convergence of finite-size effects, and the ability to
explore a wide range of parameters as $T \rightarrow 0$. We found
two important effects in the electronic LDOS: ({\em i\/}) a
suppression in the strength of the zero-bias peak at the vortex
core, and ({\em ii\/}) the appearance of satellite peaks at
frequencies of order $\pm \omega_v$, where $\omega_v$ is the vortex
oscillation frequency.

An extension of our results to $d$-wave superconductors appears in
paper II. There we will introduce a model designed specifically to
study the influence of the low energy quasiparticles, and present
implications for experiments on the cuprate supercondutors.

\acknowledgments

We thank J.~Hoffman, P.~Nikoli\'c, and Z.~Te\v sanovi\'c for useful
discussions. This research was supported by the NSF under grant
DMR-0537077 and the Deutsche Forschungsgemeinschaft under grant BA
2263/1-1 (L.~B.).

\appendix

\section{Bogoliubov-de Gennes equations for a $d$-wave superconductor}
\label{Appendix:gauge}

In this appendix we derive local Bogoliubov-de Gennes equations for
a superconductor with $d$-wave symmetry. These are applied to the
vortex zero point motion, as in Section~\ref{sec:gen}, while complete
numerical results are deferred to paper II.

As a starting point we use the following generalized Bogoliubov-de Gennes equations, 
\begin{equation}
  \left(
    \begin{array}{cc}
      \hat{\mathcal{H}}_0 & \hat \Delta \\
      \hat \Delta^{\ast} & -\hat{\mathcal{H}}_0^{\ast}
    \end{array}
    \right)
  \left(
    \begin{array}{c}
      u_{\ell}({\bf r}) \\ v_{\ell}({\bf r})
      \end{array}
    \right)
    = \epsilon_{\ell}
  \left(
    \begin{array}{c}
      u_{\ell}({\bf r}) \\ v_{\ell}({\bf r})
      \end{array}
    \right) \;.
    \label{eq:BdG}
\end{equation}
Here,
\begin{equation}
  \hat{\mathcal{H}}_0 = \frac{1}{2 m_e} \left(-i\partial_{\bf r} + {\bf A}({\bf r}) \right)^2 - E_F
  \label{eq:Hs}
\end{equation}
is the effective single particle Hamiltonian, ${\bf A}({\bf r})$ is
the vector potential, and the gap operator $\hat\Delta =
\hat\Delta({\bf r})$ is defined by
\begin{equation}
  \label{eq:Deltaint}
  \hat \Delta({\bf r}) v({\bf r}) \equiv \int d^2 r' \,
  \Delta({\bf r} - {\bf r'}/2,{\bf r'})\, v({\bf r} - {\bf r'}) \;,
\end{equation}
where $\Delta({\bf r},{\bf r'})$ is the pair potential which is a
function of the center of mass coordinate ${\bf r}$ and the relative
coordinate ${\bf r'}$. More explicitly, $\hat\Delta$ can be written
as
\begin{equation}
  \hat\Delta({\bf r}) = \int d^2 r' \,
  \Delta({\bf r} - {\bf r'}/2,{\bf r'}) e^{-{\bf r'} \cdot \partial_{\bf r}} \;.
  \label{eq:Delta}
\end{equation}
As can easily be checked, the Bogoliubov-de Gennes equations
(\ref{eq:BdG}) are invariant under the following gauge
transformation:
\begin{align}
  u({\bf r}) & \to e^{i\chi({\bf r})} u({\bf r}) \;,\nonumber \\
  v({\bf r}) & \to e^{-i\chi({\bf r})} v({\bf r}) \;,\nonumber \\
  \Delta({\bf r},{\bf r'}) & \to e^{i\chi({\bf r}+{\bf r'}/2)+i\chi({\bf r}-{\bf r'}/2)} \Delta({\bf r},{\bf r'}) \;,\nonumber \\
  {\bf A}({\bf r}) & \to {\bf A}({\bf r}) - \partial_{\bf r} \chi({\bf r}) \;.
  \label{eq:gaugetrafo}
\end{align}
All of the above is quite general and the $s$-wave case is easily
recovered by setting $\Delta({\bf r},{\bf r'})=\Delta({\bf r})
\delta({\bf r'})$.

For a $d$-wave superconductor 
we express the relative coordinate ${\bf r'}$ in polar coordinates
$r'$ and $\theta'$. Neglecting higher harmonics we write the
$d_{x^2-y^2}$-wave order parameter as
\begin{equation}
  \Delta({\bf r},{\bf r'}) = 2\Delta({\bf r},r') \cos(2\theta') \;.
  \label{eq:dwave}
\end{equation}
It is important to note that Eq.~(\ref{eq:dwave}) already implies
some choice of gauge.

To obtain local Bogoliubov-de Gennes equations we use this order
parameter and expand the rhs of Eq.~(\ref{eq:Delta}) up to second
order in ${\bf r'}$ which certainly is a good approximation if the
dominant contribution to $\Delta({\bf r},r')$ comes from small $r'$
which we will assume here. Within this approximation, it is
straightforward to show that
\begin{equation}
  \hat\Delta = \frac{(\partial_x^2-\partial_y^2) \Delta}{4 k_F^2}
  + \frac{\partial_x\Delta\partial_x - \partial_y\Delta\partial_y}{k_F^2}
  + \frac{\Delta (\partial_x^2-\partial_y^2)}{k_F^2} \;,
  \label{eq:Delta2}
\end{equation}
which can also be written as
\begin{equation}
  \hat\Delta = \frac{\{\partial_x,\{\partial_x,\Delta({\bf r})\}\}}{k_F^2}
  - \frac{\{\partial_y,\{\partial_y,\Delta({\bf r})\}\}}{k_F^2}
  \label{eq:DeltaSL2}
\end{equation}
where $\{a,b\}=(ab+ba)/2$ denotes a symmetrized product and we have
defined $\Delta({\bf r})$ as
\begin{equation}
  \Delta({\bf r}) \equiv \frac{\pi}{4} k_F^2 \int_0^{\infty} dr'\,r'^3 \Delta({\bf r},r') \;.
  \label{eq:DeltaR}
\end{equation}
Eq.~(\ref{eq:DeltaSL2}) is equivalent to the gap operator derived by
Simon and Lee \cite{Simon97} for a $d_{xy}$-wave superconductor. It
was later claimed by Vafek {\em et al.} \cite{Vafek01} that this gap
operator would not preserve the gauge invariance of the
Bogoliubov-de Gennes equations and an additional term was introduced
to fix this problem.
The implicit assumption underlying this claim is that $\Delta({\bf
r})$ (as well as $\hat \Delta$) should transform under a gauge
transformation as $\Delta({\bf r}) \to e^{2i\chi({\bf r})}
\Delta({\bf r})$. Using our approach, it is however easy to see that
although the gap operator $\hat \Delta$ is gauge invariant the above
assumption does not hold: Working to the same order as before it
follows from Eq.~(\ref{eq:gaugetrafo}) that under a gauge
transformation we have $\Delta({\bf r},{\bf r'}) \to e^{2i\chi({\bf
r})}(1 + i \partial_{\alpha}\partial_{\beta}\chi r'_{\alpha}
r'_{\beta}/4) \Delta({\bf r},{\bf r'})$. Plugging this into
Eq.~(\ref{eq:Delta2}), it is a straightforward exercise to show that
under this transformation the gap operator $\hat \Delta$ does indeed
transform as $\hat \Delta \to e^{2i\chi({\bf r})} \hat \Delta$ such
that the local Bogoliubov-de Gennes equations are gauge invariant.
The transformation properties of $\Delta({\bf r})$ are more
complicated but this is no problem because the gap operator is a
combination of derivatives and $\Delta({\bf r})$ and only this
combination has to transform as $\hat \Delta \to e^{2i\chi({\bf r})}
\hat \Delta$. It should also be noted that the derivation of the gap
operator and the above argument about its gauge transformation
become exact in the limit where $\Delta({\bf r},r')$ becomes more
and more peaked at smaller and smaller $r'$.

\section{Effective action of a vortex in a BCS superconductor}
\label{app:vortexaction}

In this appendix we present a simple and straightforward derivation
of the effective action describing vortex dynamics in a clean
two-dimensional BCS superconductor. These complement the results of
Ref.~\onlinecite{predrag}, which presented the corresponding results
using a low energy theory for a $d$-wave superconductor.

Our starting point is the Bogoliubov-de Gennes-like action given in
Eq.~(\ref{eq:Sinit}) without the harmonic oscillator potential
$V({\bf R})$,
\begin{equation}
  \mathcal{S} = \int d^2 r\, d\tau\,(\bar \psi_{\uparrow},\psi_{\downarrow})
  \left(\partial_{\tau} +         \mathcal{H}_{\text{BdG}} \right)
  \left(\begin{array}{c}
      \psi_{\uparrow} \\
      \bar \psi_{\downarrow}
    \end{array} \right) \;.
  \label{eq:SBdG0V0}
\end{equation}
In contrast to alternative derivations of an effective vortex action
we have seen, we will not expand $\mathcal{H}_{\text{BdG}}$ in
powers of ${\bf R}$. Instead, we diagonalize the Bogoliubov-de
Gennes Hamiltonian $\mathcal{H}_{\text{BdG}}$ at every instant in
imaginary time $\tau$ in terms of its eigenfunctions. Using the
unitary transformation given in Eq.~(\ref{eq:psitochi}) we then
obtain
\begin{equation}
  \mathcal{S} = \int_0^{\beta} d\tau\, \sum_\ell \bar \chi_\ell (\partial_{\tau} + \epsilon_\ell) \chi_\ell
  + \int_0^{\beta} d\tau\, \sum_{\ell,\ell'} \bar \chi_{\ell} Q_{\ell,\ell'} \chi_{\ell'}
\;.
  \label{eq:SBdGV1}
\end{equation}
Here, $Q_{\ell,\ell'} \equiv Q_{\ell,\ell'}(\tau)$ is given by
\begin{align}
  \label{eq:Qtau}
  Q_{\ell,\ell'} = - {\bf \dot R}({\tau}) \cdot & \int_{|{\bf r}| \le R_0} \!\!\!\! d^2 r\, \big[ u_{\ell}^{\ast}({\bf r} - {\bf R}(\tau)) \partial_{{\bf r}} u_{\ell'}({\bf r} - {\bf R}(\tau)) \nonumber \\
& {} + v_{\ell}^{\ast}({\bf r} - {\bf R}(\tau)) \partial_{{\bf r}}
v_{\ell'}({\bf r} - {\bf R}(\tau))\big] \;.
\end{align}
When shifting ${\bf r}$ by ${\bf R}(\tau)$, boundary terms which
finally will lead to a Berry phase need to be considered carefully.
We have therefore restricted the integration over space to $|{\bf
r}| \le R_0$ and will only at the end of our calculation take the
limit $R_0 \to \infty$. Doing the above mentioned shift ${\bf r} \to
{\bf r} + {\bf R}(\tau)$, we get $Q_{\ell,\ell'}(\tau) =
Q_{\ell,\ell'}^{(0)}(\tau) + Q_{\ell,\ell'}^{(1)}(\tau)$ with
\begin{equation}
  \label{eq:Q0}
    Q_{\ell,\ell'}^{(0)}(\tau) = - {\bf \dot R} \cdot \int_{|{\bf r}| \le R_0} d^2 r\, \big[ u_{\ell}^{\ast} \partial_{{\bf r}} u_{\ell'} + v_{\ell}^{\ast} \partial_{{\bf r}} v_{\ell'} \big] \;,
\end{equation}
and
\begin{equation}
  \label{eq:Q1}
  Q_{\ell,\ell'}^{(1)}(\tau) = \oint_{|{\bf r}| = R_0} \!\!\!\! (d{\bf r} \times {\bf R}) \cdot {\bf \hat e}_z \left( {\bf \dot R} \cdot \big[ u_{\ell}^{\ast} \partial_{{\bf r}} u_{\ell'} + v_{\ell}^{\ast} \partial_{{\bf r}} v_{\ell'} \big] \right) \;.
\end{equation}
To obtain an effective action for the vortex, we can now integrate
out the fermionic degrees of freedom. It is convenient to first
transform the imaginary-time integral in Eq.~(\ref{eq:SBdGV1}) in a
Matsubara sum over fermionic frequencies $\tilde \omega_n =
(2n+1)/\beta$. The effective action for the vortex is then given by
\begin{equation}
  \mathcal{S}_{\text{eff}}^{\text{Vortex}} =
  - \text{Tr} \ln\left(\openone + \frac{1}{\beta} {\bf G Q}\right)
  \;.
  \label{eq:SeffVortex}
\end{equation}
Here, ${\bf G}$ is a diagonal matrix Green function with matrix
elements $1/(i\tilde \omega_n - \epsilon_{\ell})$, ${\bf Q}$ has as
its matrix elements the Fourier transforms of
$Q_{\ell,\ell'}(\tau)$, and the trace is over both $\tilde \omega_n$
and all $\ell$. Expanding the logarithm in the effective vortex
action up to second order we obtain
$\mathcal{S}_{\text{eff}}^{\text{Vortex}} \approx
\mathcal{S}_{\text{eff},1}^{\text{Vortex}} +
\mathcal{S}_{\text{eff},2}^{\text{Vortex}}$, with
\begin{equation}
  \label{eq:SeffVortex1}
  \mathcal{S}_{\text{eff},1}^{\text{Vortex}}
  = \sum_{\ell} f(\epsilon_{\ell}) {\bf Q}_{\ell,\ell}^{(1)}(0) \;,
\end{equation}
and
\begin{equation}
  \label{eq:SeffVortex2}
  \mathcal{S}_{\text{eff},2}^{\text{Vortex}}
  = \frac{1}{2}\,\frac{1}{\beta} \sum_{\omega_m,\ell,\ell'} \frac{f(\epsilon_{\ell'}) -f(\epsilon_{\ell})}{i\omega_m - \epsilon_{\ell} + \epsilon_{\ell'}}\, |{\bf Q}_{\ell,\ell'}^{(0)}(i\omega_m)|^2 \;.
\end{equation}
Here, $f(\epsilon) = 1/(e^{\beta \epsilon} + 1)$ is the Fermi
function and the $\omega_m = 2\pi m/\beta$ are bosonic Matsubara
frequencies. While ${\bf Q}_{\ell,\ell'}^{(0)}$ does not give a
contribution to Eq.~(\ref{eq:SeffVortex1}) due to its antisymmetry,
we have not included ${\bf Q}_{\ell,\ell'}^{(1)}(i\omega_m)$ into
Eq.~(\ref{eq:SeffVortex2}) because it only contributes at higher
order. Identifying
\begin{equation}
  \label{eq:jr}
  {\bf j}({\bf r}) = \frac{1}{2 m_e i} \sum_{\ell}  \big[ u_{\ell}^{\ast} \partial_{{\bf r}} u_{\ell'} + v_{\ell}^{\ast} \partial_{{\bf r}} v_{\ell'} \big]
\end{equation}
as the current and using well-known vector identities we can rewrite
$\mathcal{S}_{\text{eff},1}^{\text{Vortex}}$ as
\begin{align}
  \label{eq:SeffVortex1j}
  \mathcal{S}_{\text{eff},1}^{\text{Vortex}}
  = \frac{2 m_e i}{\beta} \sum_{\omega_m} & i\omega_m
  \Big( {\bf R}(i\omega_m) \times
    \Big[ {\bf R}(-i\omega_m) \oint d{\bf r \cdot j} \nonumber \\
 & {}- \oint d{\bf r}\, ({\bf R}(-i\omega_m) {\bf \cdot j}) \Big] \Big) \cdot {\bf \hat e}_z \;.
\end{align}
Now, using ${\bf j} = \rho_s {\bf v}_s = \rho_s {\bf \hat
e}_{\theta}/2 m_e r$, where $\rho_s$ is the superfluid stiffness and
${\bf v}_s$ the superfluid velocity, we arrive at
\begin{equation}
  \label{eq:SeffVortex1jII}
  \mathcal{S}_{\text{eff},1}^{\text{Vortex}}
  = - \pi \rho_s \frac{1}{\beta} \sum_{\omega_m} \omega_m
  \left({\bf R}(i\omega_m) \times {\bf R}(-i\omega_m) \right) \cdot {\bf \hat e}_z \;.
\end{equation}
This is the Magnus force whose interpretation as a geometrical Berry
phase is most transparent when transforming back to imaginary time.
Defining
\begin{equation}
  \label{eq:SGamma}
  S(\Gamma) = \frac{1}{2} \int_0^{\beta} d\tau\, \left( {\bf \dot R} \times {\bf R} \right) \cdot {\bf \hat e}_z
  = \frac{1}{2} \int_{\Gamma} \left( d{\bf R} \times {\bf R} \right) \cdot {\bf \hat e}_z
\end{equation}
as the area enclosed by the loop $\Gamma$ we can recast
Eq.~(\ref{eq:SeffVortex1jII}) into
\begin{equation}
  \label{eq:SeffVortex1SG}
  \mathcal{S}_{\text{eff},1}^{\text{Vortex}}
  = - i 2 \pi \rho_s S(\Gamma) \;,
\end{equation}
which agrees with the result by Ao and Thouless who emphasized the
robustness of the Berry phase.\cite{Ao93}
Let us now consider $\mathcal{S}_{\text{eff},2}^{\text{Vortex}}$
which we rewrite as
\begin{equation}
  \label{eq:SeffVortex2WR}
  \mathcal{S}_{\text{eff},2}^{\text{Vortex}}
  = \frac{1}{2 \beta} \!\! \sum_{\omega_m,\ell,\ell'} \!\! \frac{f(\epsilon_{\ell'}) -f(\epsilon_{\ell})}{i\omega_m - \epsilon_{\ell} + \epsilon_{\ell'}}\,\omega_m^2\,|{\bf U}_{\ell,\ell'} \cdot {\bf R}(i\omega_m)|^2 \;,
\end{equation}
with
\begin{equation}
  \label{eq:W}
  {\bf U}_{\ell,\ell'} = \int 
  d^2 r\, \big[ u_{\ell}^{\ast} \partial_{{\bf r}} u_{\ell'} + v_{\ell}^{\ast} \partial_{{\bf r}} v_{\ell'} \big] \;.
\end{equation}
Neglecting the small $i \omega_m$ in the denominator of
Eq.~(\ref{eq:SeffVortex2WR}) and noticing that after summation over
$\ell$ and $\ell'$ the term proportional to $({\bf R}(i\omega_m)
\times {\bf R}(-i\omega_m)) \cdot {\bf \hat e}_z$ has to vanish we
obtain
\begin{equation}
  \label{eq:SeffVortex2M}
  \mathcal{S}_{\text{eff},2}^{\text{Vortex}}
  = \frac{m_v}{2}\,\frac{1}{\beta} \sum_{\omega_m} \omega_m^2\,|{\bf R}(i\omega_m)|^2
  = \frac{m_v}{2} \int_0^{\beta} d\tau\, {\bf \dot R}^2(\tau)
 \;,
\end{equation}
where the mass of the vortex, $m_v$, is given by
\begin{equation}
  \label{eq:mv}
  m_v = \frac{1}{2} \sum_{\omega_m,\ell,\ell'} \frac{f(\epsilon_{\ell}) -f(\epsilon_{\ell'})}{\epsilon_{\ell} - \epsilon_{\ell'}}\,|{\bf U}_{\ell,\ell'}|^2 \;.
\end{equation}
Using the formalism described in the main body of this paper this
equation can be used to calculate the mass of a vortex in a
superconductor with $s$- or $d$-wave symmetry.

To summarize, we can write the vortex action as
\begin{equation}
  \mathcal{S}_{\text{eff}}^{\text{Vortex}} = \frac{m_v}{2}\, \frac{1}{\beta} \sum_{\omega_m} \,
  {\bf R}^{\dagger} (i\omega_m)
  \left( \begin{array}{cc}
      \omega_m^2 & \omega_c \omega_m \\
      -  \omega_c \omega_m & \omega_m^2
    \end{array} \right) {\bf R} (i\omega_m) \;,
\end{equation}
(with $\omega_c = 2\pi\rho_s/m_v$) which when taking the limit of
zero temperature and putting the vortex in a harmonic oszillator
potential turns into Eq.~(\ref{eq:Vortex}). While the vortex mass
can in principle be calculated microscopically it can also be
treated as a phenomenological constant.

\end{document}